\documentclass[draftcls,onecolumn,peerreview]{IEEEtran}
\usepackage{amsmath}
\usepackage{amssymb}
\usepackage{amsbsy}
\usepackage[dvips]{graphicx}

\newtheorem{cor}{Corollary}

\newtheorem{prop}{Proposition}
\newtheorem{rem}{Remark}
\newtheorem{lem}{Lemma}
\newcommand{\eps}{\epsilon}
\newcommand{\defeq}{\triangleq}

\newcommand{\be}{\begin{equation}}
\newcommand{\ee}{\end{equation}}
\newcommand{\bea}{\begin{eqnarray}}
\newcommand{\eea}{\end{eqnarray}}

\newcommand{\ben}{\begin{enumerate}}
\newcommand{\een}{\end{enumerate}}
\newcommand{\bit}{\begin{itemize}}
\newcommand{\eit}{\end{itemize}}
\newcommand{\bpf}{\begin{proof}}
\newcommand{\epf}{\end{proof}}

\begin{document}
\title{Scaling Laws and Techniques in Decentralized Processing of Interfered Gaussian Channels}

\author{\authorblockN{Amichai Sanderovich, Michael Peleg and Shlomo Shamai (Shitz)}\\
\authorblockA{Technion, Haifa, Israel \\
Email: \{amichi@tx,michael@ee,sshlomo@ee\}.technion.ac.il}
 }
\maketitle
\begin{abstract}
The scaling laws of the achievable communication rates and the
corresponding upper bounds of distributed reception in the
presence of an interfering signal are investigated. The scheme
includes one transmitter communicating to a remote destination via
two relays, which forward messages to the remote destination
through reliable links with finite capacities. The relays receive
the transmission along with some unknown interference. We focus on
three common settings for distributed reception, wherein the
scaling laws of the capacity (the pre-log as the power of the
transmitter and the interference are taken to infinity) are
completely characterized. It is shown in most cases that in order
to overcome the interference, a definite amount of information
about the interference needs to be forwarded along with the
desired message, to the destination. It is exemplified in one
scenario that the cut-set upper bound is strictly loose. The
results are derived using the cut-set along with a new bounding
technique, which relies on multi letter expressions. Furthermore,
lattices are found to be a useful communication technique in this
setting, and are used to characterize the scaling laws of
achievable rates.
\end{abstract}
\section{Introduction}
In this paper, we treat the problem of decentralized detection, with
an interfering signal. Decentralized detection \cite{fullpaper} is
an interesting and timely setting, with many applications such as
the emerging 4G networks \cite{full_oren}, \cite{Conf_ISIT2007Oren},
smart-dust and remote inference to name just a few.

This setting consists of a transmitter which communicates to a
distant destination via intermediate relay/s which facilitate the
communication, where no direct link between the transmitter and
destination is provided. The model includes reliable links with
fixed capacities between the relays and the destination. Such
model is further extended in \cite{fadingfull} to incorporate a
fading channel between the transmitter and the relays.

The setting also includes an interference, which is modeled as a
Gaussian white signal (no encoding is assumed). The interfering
signal is unknown to either the transmitter, the destination or
the relays. Such setting suits numerous real-world scenarios such
as airport tower communication which need to have more than one
reception point for increased security against jamming, hot-spots
operating in a dense interference environment and cellular network
with a strong interference.

This model is somewhat different than that treating the jamming
problem as a minmax optimization, where the jammer is optimized to
block the communication of the transmitter, which in turn is
optimized to maximize the reliably conveyed rate \cite{Hedby1994},
\cite{TekinYener2007}. Some recent papers that deal with similar
yet different settings are
\cite{MaricDaboraGoldsmith2008_I},\cite{MaricDaboraGoldsmith2008_II}
and \cite{DaboraMaricGoldsmith2008} and also
\cite{SahinSimeoneErkip2009}. The concept of \textit{generalized
degrees of freedom}, used in \cite{WangTse2009} is intimately
related to the scaling laws defined in this paper. The only remedy
offered in this paper is the exploitation of the spatial
correlation, where different relays are receiving the same jamming
signal. In order to efficiently overcome the jammer in a
distributed manner, we use lattice codes, which enable efficient
modulo-like operation, filtering out some of the undesired
interference.

Lately, several contributions suggested scenarios in which
structured codes, and lattices in particular demonstrated to
outperform best known random coding strategies
\cite{NazerGastpar2007},\cite{NazerGastpar2007_I}
\cite{KrithivasanPradhan2007} and
\cite{PhilosofKhistiErezZamir2007}. These works apply the new
advances in the understanding of lattice codes
\cite{Poltyrev1994},\cite{ErezZamir2004} and
\cite{ZamirShamaiErez2002}, and nested lattices
\cite{ForneyTrottChung2000}, \cite{ConwaySloane1982}, and
specifically their ability to perform well as both source and
channel codes. Specifically, \cite{NazerGastpar2008} used lattice
codes for interference channel.

Additional works relevant in this respect are
\cite{MallikKotter2008} which uses lattices to overcome known
interference (by a remote helper node), and is a special case of
\cite{PhilosofZamir2008}, which discusses the capacity of the
doubly dirty MAC channel. These two works deal with a helper node
which aids the transmitter to overcome an interference. Although
there the interference is known to the helper node, this is still
similar to our model, where instead of a channel, the helper node
has a reliable link with finite capacity to the destination.

Several other works are relevant here. The capacity for distributed
computation of the modulo function with binary symmetric variables
is given in \cite{KronerMarton1976}, while distributed computation
with more general assumptions is provided in
\cite{OrlitskyRoche2001}. A deterministic approach to wireless network, which specifically fits the scaling law measure, is given in \cite{AvestimehrDiggaviTse2007}.

We will focus on three main scenarios, each one models different
constraints imposed on the resources available for communication.
Each of these models reveals another aspect of the general
problem, while their joint contribution is demonstrating the same
principles.

The scaling laws derived for these scenarios reveal that a
definite amount of information about the interference is required
at the destination if reliable communication is established. This
conclusion is an extension of a special case of the results of
Kr\"{o}ner and Marton \cite{KronerMarton1976}, for two separated
\textit{independent} binary sources $X,Y$, where a remote
destination wishes to retrieve $X\oplus Y$, and needs to get both
$X,Y$. The difference being that in \cite{KronerMarton1976},
random coding strategies sufficed to demonstrate this principle,
while here we resort to lattice codes.

Further, for dependent binary sources $X,Y$, \cite{KronerMarton1976}
demonstrated the advantages of lattice codes over random codes necessitating the knowledge
of both $(X,Y)$.

All the achievable rates in this paper are derived, focusing on
scaling laws. These are by no means optimal in any other sense.
Each rate expression was given for both the case of
$P_X>P_J$ and $P_X<P_J$, by using a simple minimum operation, for
simplicity and brevity ($P_X$ being the average power of the
transmitter and $P_J$ being the average power of the
interference).

This paper is divided into four sections, the first section
contains the general setting and the basic definitions, the second
presents all central results along with some discussion, the third
section contains proofs of the necessary and sufficient conditions. We
then conclude with some final notes. Some proofs are relegated to the appendices.

\section{Model and Definitions}
We denote by $X_i$ the random variable at the i-th position and by
$\boldsymbol{X}$ the vector of random variables $(X_1,\ldots,X_n)$.

\begin{figure}
\centering
 \includegraphics[width=0.75\textwidth]{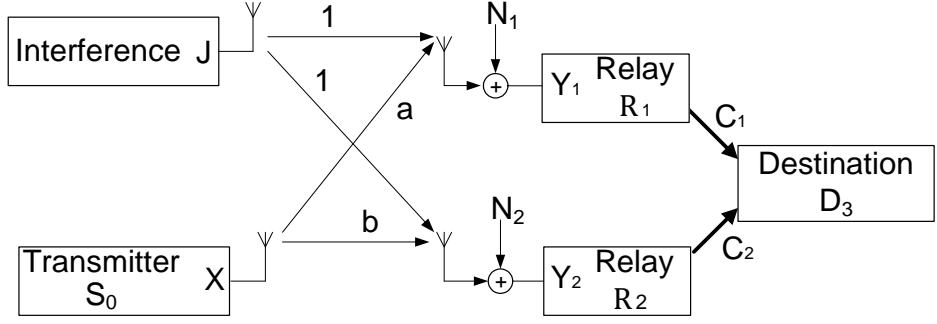}
 \caption{The system setting, with transmitter $S_0$ received by $\mathcal{R}_1$ and $\mathcal{R}_2$ with channel transfer coefficients $a$ and $b$, respectively.
 Lossless links with respective capacities of $C_1$ and $C_2$ between relays $\mathcal{R}_1,\mathcal{R}_2$ and destination $\mathcal{D}_3$.}
 \label{fig:setting}
\end{figure}

We consider the channel as it appears in Figure \ref{fig:setting},
where a source $S_0$ wishes to transmit the message $W\in
\{1,\dots,2^{nR}\}$ to the destination $D_3$ by transmitting
$\boldsymbol{X}$ to the channel, where $R$ designates the
communication rate. The transmitter is limited by an average power
constrain $\sum_{k=1}^n \mathrm{E}[X_k^2]\leq P_X$. The channel has
two outputs, $Y_1$ and $Y_2$, where
\bea \label{eq:def1}Y_1 &=& a X + J + N_1\\
Y_2&=& b X + J + N_2.\label{eq:def2} \eea
The additives $J,N_1,N_2$ are Gaussian
memoryless independent processes, with zero mean and variances of
$P_J,P_{N_1},P_{N_2}$, respectively, and $a,b\in\mathbb{R}$ are
two fixed coefficients to be addressed later. The two channel
outputs $Y_1,Y_2$ are received by two distinct relay units,
$\mathcal{R}_1$ and $\mathcal{R}_2$, respectively. These relays
use separate processing on the received signals $Y_1$ and $Y_2$,
and then forward the resultant signal to the destination $D_3$. It
is noted that the destination is only interested in the message
$W$. The relays can forward messages to the destination over
reliable links with finite capacities of $C_1$ and $C_2$ bits per
channel use. The destination then decides on the transmitted
message $\hat{W}$. A communication rate $R$ is said to be
achievable, if the average error probability at the destination
$\Pr\{W\neq \hat{W}\}$ is arbitrarily close to zero for
sufficiently large $n$.

The communication system is therefore completely characterized by
four deterministic functions ($I_n\triangleq[1,\dots,2^n]$) \bea \boldsymbol{X}(W)=\phi_{S_0}(W)&:&I_{nR}\rightarrow \mathbb{R}^n\\
V_1(\boldsymbol{Y}_1)=\phi_{\mathcal{R}_1}(\boldsymbol{Y}_1)&:&\mathbb{R}^n\rightarrow I_{nC_1}\\
V_2(\boldsymbol{Y}_2)=\phi_{\mathcal{R}_2}(\boldsymbol{Y}_2)&:&\mathbb{R}^n\rightarrow I_{nC_2}\\
\hat{W}(V_1,V_2)=\phi_{D_3}(V_1,V_2)&:&I_{nC_1}\times
I_{nC_2}\rightarrow I_{nR}. \eea

We can divide the general case of (\ref{eq:def1})-(\ref{eq:def2}) into three possible options:
\begin{enumerate}
\item $a=0$ or $b=0$,\ $a\neq b$.
\item $a,b\neq 0,\ a\neq b$.
\item $a=b$.
\end{enumerate}
The last case is of no real interest since the scaling behavior is obvious and remains the same for one or two relays, so this paper will focus only on the first two.
Similarly, for $C_1,C_2$ we have three cases, where $C\rightarrow\infty$ means that $C$ goes to infinity much faster than $P_X,P_J$:
\begin{enumerate}
\item $C_1\rightarrow\infty$ or $C_2\rightarrow\infty$.
\item Neither $C_1\rightarrow\infty$ nor $C_2\rightarrow\infty$.
\item $C_1\rightarrow\infty$ and $C_2\rightarrow\infty$.
\end{enumerate}
The last case results with full cooperation. In this case the achievable rate and the upper bound are identical and equal to the Shannon capacity,
given by (for example for $a=1,b=-1$):
\begin{equation}
R=\frac{1}{2}\log_2(1+2P_X).
\end{equation}
This rate is achieved by using maximal ratio combining of the two receptions, completely eliminating the interference.
So the general case in this paper reduces to three main scenarios, in which we investigate
the scaling laws of $C_1$ and $C_2$, as a function of the
achievable rate $R$. These three cases consist of the four possible options described above for $a,b,C_1,C_2$, where the option of $C_1\rightarrow\infty$ or $C_1\rightarrow\infty$ while $a b\neq 0$ was dropped since it is very similar to the case where either $C_1\rightarrow\infty$ or $C_1\rightarrow\infty$ when $a=0$ or $b=0$ while $a\neq b$.

Since the channel between the transmitter and
the relays can support a rate with scaling of up to
$\lim_{P_X\rightarrow\infty}\frac{R}{\log_2(P_X)}=\frac{1}{2}$, we
investigate the required capacities of the links to achieve this
scaling, and also the degradation of $R$ when they are smaller.

We denote by scaling the pre-log coefficient defined by the limit
$scaling=\lim_{P_X\rightarrow\infty}\frac{R}{\log_2(P_X)}$, and
write it for the sake of brevity as $R\sim scaling\ \log_2(P_X)$.
Similarly, the relation
$\lim_{P_X\rightarrow\infty}\frac{R}{\log_2(P_X)}\geq \frac{1}{2}$
is designated by $R\gtrsim \frac{1}{2}\log_2(P_X)$.

\subsection*{Case A: Relaying the Interference}
The scenario here specializes to \bea a&=&1\\
b&=&0\\
P_{N_1}&=&1\\
P_{N_2}&=&0\\
C_1&\rightarrow&\infty.
 \eea
The last condition simply states that the destination receives the
channel output, which is composed of the transmission plus some
unknown interference, which may degrade or even prevent any reliable
decoding. The interference is received intact in $\mathcal{R}_2$,
and then relayed to the destination to enable reliable decoding.
We consider a fixed $a=1$ for simplicity. The results for any $a\neq 0$ are obtained from the results for the fixed $a=1$ almost verbatim.

This model can describe a situation where an additive jammer is
known to a relay which can assist the destination with resolving
the transmission from $S_0$. Notice that for an infinite Jammer
power ($\frac{P_J}{P_X}\rightarrow\infty$), a link to the relay
$\mathcal{R}_2$ is necessary to achieve any positive rate.

\subsection*{Case B: Relaying the Interfered Signal and the Interference}
The only change compared to the previous Scenario A is that here
$C_1$ is finite. The added limitation extends the decentralization
inherent in the scheme, which models practical systems where the
destination is not collocated near either relays.

\subsection*{Case C: Relaying 2 Interfered Signals}
In this scenario we consider the case where
\bea a&=&1\\
b&=&-1\\
P_{N_1}&=&1\\
P_{N_2}&=&1.
 \eea
In this case, the signals are in anti-phase and hence joint
processing ($C_1,C_2\rightarrow\infty$) via substraction $Y_1-Y_2$
would completely remove the interference, and would allow for
reliable rate of $R=\frac{1}{2}\log_2(1+2P_X)$.\\
As in the previous cases, the solution to $a=1,b=-1$, gives the same scaling behavior as taking any $a\neq b,\ a\neq 0,\ b\neq 0$. Notice that for scaling, the destination still wants to cancel the interference, and thus still needs to perform $Y_1-Y_2$, regardless of the actual $a,b$, as long as $a\neq b$.
Also the finite $P_X,P_J$ results for this general case are readily derived following the same steps.
\section{Main Results}\label{sec:main}
The main results described in this Section are divided into the
three scenarios detailed above, and include both the achievable
rates and outer bounds. The resulting scaling laws of the inner and
outer bounds coincide.
\subsection*{Case A: Relaying the Interference}\label{subsec:R_I_S}
\begin{prop}\label{prop:R_I_B}\textit{
\begin{enumerate}
\item\label{subprop:R_I_Binner}The following rate for Case A
is achievable
\begin{equation}\label{eq:rate_1}
R\leq\max\left\{\frac{1}{2}\log_2
\bigg(\frac{1+P_X}{1+\min\left\{1+P_X,\frac{P_JP_X}{P_X+1}\right\}2^{-2C_2}}
\bigg),0\right\}.
\end{equation}
\item\label{subprop:R_I_Bouter}An upper bound for all achievable rates for Case A is (cut-set bound)\be\label{eq:A_cutset} R\leq \min\left\{C_2+I(X;Y_1),
I(X;Y_1|J)\right\} \ee which for the Gaussian channel reads,
\be\label{eq:caseA_B} R\leq
\min\left\{C_2+\frac{1}{2}\log_2\left(1+\frac{P_X}{P_J+1}\right),
\frac{1}{2}\log_2(1+P_X)\right\}. \ee 
\end{enumerate}}
\end{prop}
The proof of Proposition \ref{prop:R_I_B} appears in Section \ref{sec:out_proof}, where the achievable rate is just a special case of the achievable rate of Case B. It is understood that the
achievable rate (\ref{eq:rate_1}) is not optimal, but it does
prove the scaling laws.

From Proposition
\ref{prop:R_I_B} the scaling laws can be derived.
\begin{prop}\label{prop:scaling_simple}
\textit{To achieve a scaling of $R\sim\frac{1}{2}\log_2(P_X)$, when
relaying the interferer (Case A), the sufficient and necessary
lossless link capacity scaling is \be\label{eq:up_A}
C_2\gtrsim\frac{1}{2}\log_2\left(\frac{P_X P_J}{P_X+P_J}\right). \ee
Furthermore, the gap between the achievable rate and the upper bound is no more than 1 bit.}
\end{prop}
This Proposition is proved in Appendix \ref{app:gap}.
The next corollary is a special case of Proposition
\ref{prop:scaling_simple}:
\begin{cor}
A scaling of $C_2\sim\frac{1}{2}\log_2(P_X)$ is sufficient, for any
interferer, regardless of its power and statistics.
\end{cor}
This Corollary holds, since the proof for the achievable rate in
section \ref{sec:inner_proof} uses random dithering, which
achieves the same performance for any $J$ which is independent of the transmitted signal.

For example, when the interferer is another transmitter, the
robustness is with respect to the the applied code, modulation
technique and interference power. However, the exact phase,
between the reception at $Y_1$ and $Y_2$ is still required.

\subsection*{Case B: Relaying the Interfered Signal and the Interference}
Now also $C_1$ is finite.
\begin{prop}\label{prop:R_SI_B}\textit{\begin{enumerate}
\item\label{subprop:R_SI_Binner} The following rate for Case B is achievable
\begin{equation}\label{eq:R_2_stage4}
R<\max\left\{\frac{1}{2}\log_2\left(\frac{(1+P_X)(2^{2C_1}-1)}{P_X+
2^{2C_1}+\min\left\{1+P_X,\frac{P_JP_X}{P_X+1}\right\}2^{-2C_2}(2^{2C_1}-1)}\right),0\right\}.
\end{equation}
\item\label{subprop:R_SI_Bouter}An upper bound for all achievable rates of Case B is (cut-set bound)\be\label{eq:B_cutset} R\leq \min\left\{C_1,C_2+I(X;Y_1),
I(X;Y_1|Y_2)\right\} \ee which for the underlying Gaussian channel
turns out to be \be\label{eq:B_outer} R\leq
\min\left\{C_1,C_2+\frac{1}{2}\log_2\left(1+\frac{P_X}{P_J+1}\right),\frac{1}{2}\log_2(1+P_X)\right\}.
\ee
\end{enumerate}
}
\end{prop}
The proof of Proposition \ref{prop:R_SI_B} appears in Section \ref{sec:inner_proof}. It is understood that the
achievable rate (\ref{eq:R_2_stage4}) is not optimal, but it does
prove the scaling laws.

 Next, we quantify the
necessary scaling of the link capacity $C_1$.
\begin{prop}\label{prop:relay_sig_inrfr}
\textit{To achieve a scaling of $R\sim\frac{1}{2}\log_2(P_X)$, when
relaying the interfered signal and the interference (Case B),
sufficient and necessary lossless links capacities scale as
\begin{eqnarray}
C_1&\gtrsim&\frac{1}{2}\log_2(P_X)\\
C_2&\gtrsim&\frac{1}{2}\log_2\left(\frac{P_X P_J}{P_X+P_J}\right).
\end{eqnarray}
Furthermore, the gap between the achievable rate and the upper bound is no more than 1.29 bits.}
\end{prop}
The proof appears in Appendix \ref{app:gap}.

This Proposition establishes that $C_1$, that is the capacity of the
link from the relay that receives the signal and the interference to
the destination, scales the same as if there was no interference.
\begin{cor} A scaling of $C_1\sim\frac{1}{2}\log_2(P_X)$ suffices to
achieve robustness against any interference, regardless of its power
or statistics, as long as it remains independent of $X$.
\end{cor}
In Figure \ref{fig:rate}, the scaling of the achievable rate of
Case B is drawn as a function of $C_1+C_2$ for $P_J<P_X$. For the
sake of the achievable rate $C_1,C_2$ were selected such that $C_1+C_2$ is fixed and the achievable rate is maximized. The cut-set upper bound in Case B is met by an
achievable rate along the entire range in Figure \ref{fig:rate}.
Specifically, from the point $C_1+C_2=0$ to
$C_1+C_2=\frac{1}{2}\log_2(P_X/P_J)$ ($P2$), an achievable scheme
which uses simple local decoding at $\mathcal{R}_1$ is optimal.
This is since using the decoded information rate equals the
cut-set bound ($C_1\geq R$), which is therefore tight even for
finite $P_X,P_J$. The slope of the curve is 1, since only
information bits are forwarded. Such local decoding is optimal as
long as $R\leq \frac{1}{2}\log_2(1+\frac{P_X}{P_J+1})$. Higher
sum-links-rate benefit by devoting some bandwidth also to the
message from $\mathcal{R}_2$. The achievable rate of Proposition
\ref{prop:R_SI_B}, equation (\ref{eq:R_2_stage4}) outperforms
local decoding. In such scheme both relays basically forward the
received signals to the destination, where the signals are
subtracted at the destination, which eliminates the interference.
Thus every additional forwarded information bit requires also one
bit for forwarding the interference. This means that the rate
increases only as $\frac{1}{2}(C_1+C_2)$. The outer bound for the
range between $P2$ and $C_1+C_2=\frac{1}{2}\log_2(P_X P_J)$ ($P1$)
is due to the diagonal cut-set upper bounds ($C_2\gtrsim
R-\frac{1}{2}\log_2\left(1+\frac{P_X}{P_J}\right)$). The maximal
rate is $\frac{1}{2}\log_2(P_X)$, which is reached only when
$C_1+C_2\gtrsim \frac{1}{2}\log_2(P_XP_J)$ at $P1$.
\subsection*{Case C: Relaying 2 Interfered Signals}
In this case the desired signal is received by both relays along
with the common interference.
\begin{prop}\label{prop:I_B}\textit{\begin{enumerate}
\item\label{subprop:R_I_Binner}An achievable rate for Case C is
\begin{equation}\label{eq:achiv_C}
R>\max\left\{\frac{1}{2}\log_2\left(\frac{P_X}{\frac{P_X }{P_X+
1}+\frac{P_X}{2^{2C_1}-1}+\min\{P_X,P_J\frac{P^2_X}{(P_X+1)^2}\}2^{-2C_2}}\right),0\right\}.
\end{equation}
and this holds also with the indices 1 and 2 interchanged.
\item\label{subprop:R_I_Bouter} An upper bound for all achievable rates for Case C is (cut-set bound), \be\label{eq:cutset3} R\leq
\min\left\{C_1+C_2,C_1+I(X;Y_2),C_2+I(X;Y_1), I(X;Y_1,Y_2)\right\}
\ee which for the underlying Gaussian channel turns out to be
\be\label{eq:cutset_C} R\leq
\min\left\{C_1+C_2,C_1+\frac{1}{2}\log_2\left(1+\frac{P_X}{P_J+1}\right),C_2+\frac{1}{2}\log_2\left(1+\frac{P_X}{P_J+1}\right),\frac{1}{2}\log_2(1+2P_X)\right\}.
\ee Another upper bound for all achievable rates for Case C is
(Modulo bound)
\begin{equation}\label{eq:R_imprv_B}
R\leq\frac{1}{2}\left(C_1+C_2+\frac{1}{2}\log_2\left(1+\frac{P_X}{P_J}\right)\right)+\frac{1}{4}\log_2(8\pi
e).
\end{equation}
\end{enumerate}}
\end{prop}
The proof for Proposition \ref{prop:I_B} appears in Section \ref{sec:inner_proof}. The achievable rate
(\ref{eq:achiv_C}) is not optimal, but it does prove the scaling
laws.

Note that (\ref{eq:R_imprv_B}) states an upper bound for any $R$,
including finite rates. However, the bound is interesting only in
the case of large $P_X,P_J$, because of the added 1.55
($\frac{1}{4}\log_2(8\pi e)$) bits per channel use.

\begin{prop}\label{prop:case_C}\textit{Necessary and sufficient conditions on $C_1,C_2$, to achieve the scaling of $R$ when $P_X,P_J$ are taken to
infinity are
\begin{equation}\label{eq:scaling_C}
\begin{cases}C_1+C_2\gtrsim \max\left\{2R-\frac{1}{2}\log_2\left(1+\frac{P_X}{P_J}\right),R\right\}\\
C_1,C_2\gtrsim R-\frac{1}{2}\log_2\left(1+\frac{P_X}{P_J}\right)
\end{cases}
\end{equation}}
\textit{Furthermore, the gap between the achievable rate and the outer bound in the asymptotic regime when $R\sim0.5\log_2(P_X)$ is bounded to 2.816 bits.}
\end{prop}
The proof of Proposition \ref{prop:case_C} appears in appendices \ref{app:scaling} and \ref{app:gap}.

\begin{figure}
\centering
 \includegraphics[width=0.75\textwidth]{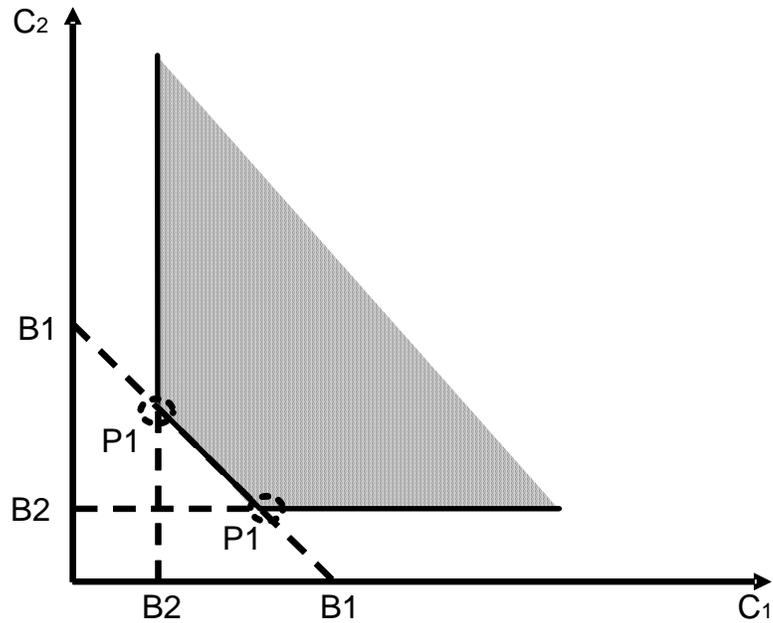}
 \caption{The scaling of $C_1,C_2$ as a function of the achievable rate $R$, where $B1$ is the scaling of $\max\left\{2R-\frac{1}{2}\log_2\left(1+\frac{P_X}{P_J}\right),R\right\}$ and $B2$ is the scaling of $R-\frac{1}{2}\log_2\left(1+\frac{P_X}{P_J}\right)$. The filled area denotes $C_1$ and $C_2$ which enable communication at rate  $R$.}
 \label{fig:region}
\end{figure}

The resulting scaling of the rate region is presented in Figure
\ref{fig:region}, where the required scaling of
$C_1,C_2$, so that the achievable rate has the scaling of $R$ is filled. The bound $B1$
in Figure \ref{fig:region} stands for the bounds on $C_1+C_2$ such
that
$C_1+C_2\gtrsim\max\left\{2R-\frac{1}{2}\log_2\left(1+\frac{P_X}{P_J}\right),R\right\}$,
while $B2$ stands for the diagonal bounds, which separately limit
$C_1,C_2$ such that $C_1,C_2\gtrsim
R-\frac{1}{2}\log_2\left(1+\frac{P_X}{P_J}\right)$. It is evident
that any increase of $C_1$ or $C_2$ can indeed only help, and the
rate-region is convex. Achieving the points $P1$ and $P2$, allows
to achieve any other point in the interior rate-region, through
time sharing.

\begin{figure}
\centering
 \includegraphics[width=0.75\textwidth]{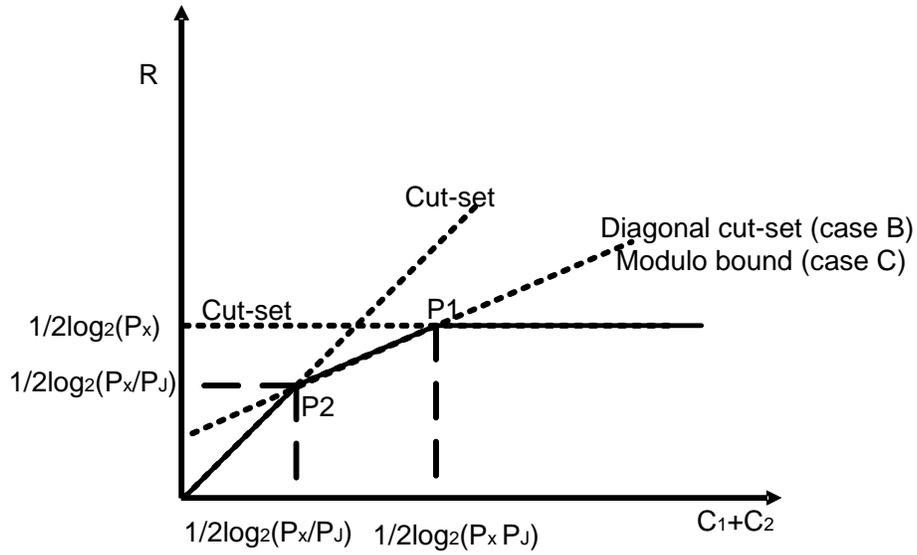}
 \caption{The achievable scaling of the rate $R$ as a function of $C_1+C_2$, for cases B and C. Upper bounds are drawn with dotted lines, while the full line is the achievable rate. It is evident that the cut-set bound in Case C is not tight, and the modulo bound was needed to characterize the scaling laws. For Case B, the cut-set bound is tight for the whole region.}
 \label{fig:rate}
\end{figure}

In Figure \ref{fig:rate}, the scaling of the achievable rate is
drawn as a function of the scaling of $C_1+C_2$, when $C_1=C_2$,
letting $P_J<P_X$. From the point $(0,0)$ to $P2$, an achievable
scheme uses simple local decoding at the relays. Since this scheme
uses all the links' bandwidth to forward only decoded information,
the cut-set bound is tight, and the slope of the curve is 1. Such
local decoding is possible as long as $R\lesssim
\frac{1}{2}\log_2(1+\frac{P_X}{P_J})$. Achieving higher rates
requires more than local decoding, and the achievable rate of
Proposition \ref{prop:I_B}, equation (\ref{eq:achiv_C}) is used.
The outer bound for this range is due to the modulo outer bound
(\ref{eq:R_imprv_B}). This scheme basically forwards the received
signals to the destination, where the signals are subtracted to
eliminate the interference. As in Case B, we get rid of the
interference only at the destination, which means that the rate
scaling increases only as the scaling of $\frac{1}{2}(C_1+C_2)$.
The maximal rate scaling is $\frac{1}{2}\log_2(P_X)$, which is
reached only when $C_1+C_2\gtrsim \frac{1}{2}\log_2(P_XP_J)$ at
$P1$. The modulo bound (\ref{eq:R_imprv_B}) determines the
behavior between the points $P2$ and $P1$.

\begin{cor}The cut-set upper bound is strictly loose for the
interference channel of Case C.
\end{cor}
\begin{proof} Take $P_J=\sqrt{P_X}$ and $C_1=C_2=\frac{1}{4}\log_2(P_X)$. Then the cut-set bound from Equation
(\ref{eq:cutset3}) for the scaling reads $R\gtrsim R_{cut}=
\frac{1}{2}\log_2(P_X)$, while the modulo bound of Equation
(\ref{eq:R_imprv_B}) for the scaling reads $R\gtrsim R_{mod}=
\frac{1}{4}\log_2(P_X)$. So we showed that
$R_{cut}>R_{mod}+\epsilon$ for some $\epsilon>0$.
\end{proof}

\begin{rem}\label{cor:jammer}
For cases A and B, when considering the outer bound due to the
underlying Gaussian channel, $R\leq \frac{1}{2}\log_2(1+P_X)$ and
combining propositions \ref{prop:scaling_simple} and
\ref{prop:relay_sig_inrfr} we get that for large $P_X,P_J$,
$\frac{1}{n}H(V_2)\gtrsim \frac{1}{2}\log_2\left(\frac{P_X
P_J}{P_X+P_J}\right)$. By adding also that
$H(V_2|\boldsymbol{J})=0$ since $V_2$ is a deterministic function
of $\boldsymbol{J}$, it follows that
\begin{equation}\label{eq:34}\frac{1}{n}I(V_2;\boldsymbol{J})\gtrsim \frac{1}{2}\log_2\left(\frac{P_X
P_J}{P_X+P_J}\right).
\end{equation}
So that in order to achieve a reliable rate
$R\sim\frac{1}{2}\log_2(P_X)$, a defined amount of information
about the interferer is required at the destination with scaling
$\frac{1}{2} \log_2 \min\{P_X,P_J\}$, for large
$P_X,P_J$.\\
%
%

\end{rem}
\section{Proofs for Basic Propositions}\label{sec:out_proof}\label{sec:inner_proof}
In this section we prove the basic propositions, not the propositions dealing with the scaling laws, which appear in appendices \ref{app:gap}-\ref{app:scaling}.
\subsection{Proofs for the Outer Bounds}
In this section we present the proofs of the necessary conditions of
the propositions in Section \ref{sec:main}.

\emph{Proof of Outer Bounds for Cases A,B and C in equations
(\ref{eq:A_cutset}),(\ref{eq:B_cutset}) and (\ref{eq:cutset3}):}
The cut-set outer bound \cite{CoverThomas} is simply the minimum
among all the communication rates between any two cuts of the
network \cite{CoverThomas} as is reflected in the three cases
under study. Let us show the cut-set for one such cut, for the
sake of conciseness, where the rest of the cuts readily follow.
Take the cut such that one set includes the transmitter with relay
$\mathcal{R}_1$ and therefore the other set includes
$\mathcal{R}_2$ and the destination. From \cite{CoverThomas},
Theorem 14.10.1: The achievable rate must be less than or equal to
$I(X^{(S)};Y^{(S^c)}|X^{(S^c)})$ for some single letter joint probability
distribution $P(X^{(S)},X^{(S^c)})$. In our setting with the
chosen cut, $X^{(S)}=(X,V_1)$, $Y^{(S^c)}=(Y_2,V_1)$ and
$X^{(S^c)}=V_2$, since the destination can not transmit anything. The underlying
channel is $P(Y^{(S^c)}|X^{(S^c)},X^{(S)})=P(Y_2,V_1|X,V_1,V_2)=P(Y_2|X)$. The right-most equality is since
$Y_2$ is not affected by $V_1$ or $V_2$, and the resulting Markov chain is $V_1-X-Y_2$.

To get the upper bound we need to maximize $I(X,V_1;Y_2,V_1|V_2)$
over $P(X,V_1,V_2)$. The mutual information $I(X,V_1;Y_2,V_1|V_2)$
is determined only by $P(X,V_1,V_2)$ and by the given channel
$P(Y_1,Y_2|X)$ such that $I(X,V_1;Y_2,V_1|V_2)=I(X,V_1;Y_2,V_1)$.
Since $V_1-X-Y_2$ is a Markov chain:
\begin{multline}
 I(X,V_1;Y_2,V_1) = H(V_1,X) - H(X|V_1,Y_2) = H(V_1|X) +
 I(X;V_1,Y_2)=\\
 = H(V_1|X) + I(Y_2;X) + H(V_1|Y_2) - H(V_1|Y_2,X)
\end{multline}
Due to the Markov chain above, $H(V_1|Y_2,X) = H(V_1|X)$. So that
\begin{equation}
R\leq I(Y_2;X) + H(V_1|Y_2) \leq I(Y_2;X) + C_1.
\end{equation}
Considering all the cut-sets, equation (\ref{eq:A_cutset}) follows
from
\begin{multline}
R\leq \min\left\{C_2+I(X;Y_1),
I(X;Y_1,Y_2)\right\}=\min\left\{C_2+I(X;Y_1),
I(X;Y_1|Y_2)+I(X;Y_2)\right\}\\=\min\left\{C_2+I(X;Y_1),
I(X;Y_1|J)\right\}
\end{multline}
where the last equality follows since $Y_2=J$. The complete proofs
for the inequalities in equations (\ref{eq:B_cutset}) and
(\ref{eq:cutset_C}) are omitted here since they are proved exactly
the same way. \hfill{\QED}

\emph{Proof of the Modulo Outer Bound in Case C (\ref{eq:R_imprv_B}):}\\
The basis of the proof is the representation of the transmitted
signal ($X$) by two components, one is an integer which is basically
known at the relays (with high probability), and the other is a heavily interfered real signal.\\
\emph{Definition:} For any $X$, $X^-\triangleq X \mod\sqrt{P_J}$ and
$X^+\triangleq
\left\lfloor\frac{X}{\sqrt{P_J}}\right\rfloor$\footnote{$\lfloor
X\rfloor$
represents the largest integer, which is no greater than $X$}.\\

Assuming, without loss of any generality, that
\begin{equation}\label{eq:assump}
H(V_1|\boldsymbol{X}^+,V_2)\leq H(V_2|\boldsymbol{X}^+,V_1).
\end{equation}
If (\ref{eq:assump}) is not satisfied, replace the indices of 1 and
2 in the following. Using Fano's inequality, where $\eps>0$ is
arbitrary small, for sufficiently large $n$, we get that
\begin{eqnarray}
nR&\leq& I(\boldsymbol{X};V_1,V_2)+n\eps\\
&=& I(\boldsymbol{X}^+;V_1,V_2)+I(\boldsymbol{X}^-;V_1,V_2|\boldsymbol{X}^+)+n\eps\\
&=& I(\boldsymbol{X}^+;V_1,V_2)+I(\boldsymbol{X}^-;V_1|\boldsymbol{X}^+,V_2)+I(\boldsymbol{X}^-;V_2|\boldsymbol{X}^+)+n\eps\\
&\leq& I(\boldsymbol{X}^+;V_1,V_2)+H(V_1|\boldsymbol{X}^+,V_2)+I(\boldsymbol{X}^-;\boldsymbol{Y}_2|\boldsymbol{X}^+)+n\eps\label{eq:forth}\\
&\leq& I(\boldsymbol{X}^+;V_1,V_2)+\frac{1}{2}[H(V_1|\boldsymbol{X}^+,V_2)+H(V_2|\boldsymbol{X}^+)]+h(\boldsymbol{Y}_2|\boldsymbol{X}^+)-h(\boldsymbol{Y}_2|\boldsymbol{X})+n\eps\label{eq:fifth}\\
&\leq&
I(\boldsymbol{X}^+;V_1,V_2)+\frac{1}{2}H(V_1,V_2|\boldsymbol{X}^+)+h(-\boldsymbol{X}^-+\boldsymbol{J}+\boldsymbol{N}_2)-h(\boldsymbol{J}+\boldsymbol{N}_2)+n\eps\label{eq:six}\\
&=& I(\boldsymbol{X}^+;V_1,V_2)+\frac{1}{2}[H(V_1,V_2)-I(\boldsymbol{X}^+;V_1,V_2)]+I(\boldsymbol{X}^-;-\boldsymbol{X}^-+\boldsymbol{J}+\boldsymbol{N}_2)+n\eps\label{eq:seven}\\
&\leq& \frac{1}{2}I(\boldsymbol{X}^+;V_1,V_2)+\frac{1}{2}H(V_1,V_2)+\frac{n}{2}\log_2\left(1+\frac{P_J}{1+P_J}\right)+n\eps\label{eq:eigth}\\
&\leq& \frac{n}{4}\log_2\left(2\pi
e\left(\frac{P_X}{P_J}+\frac{1}{12}\right)\right)+\frac{n}{2}(C_1+C_2)+\frac{n}{2}\log_2\left(1+\frac{P_J}{1+P_J}\right)+n\eps\label{eq:final}.
\end{eqnarray}
Where (\ref{eq:forth}) is since $H(V_1|\boldsymbol{X},V_2)\geq 0$,
and the data processing Lemma $V_2=\phi_{\mathcal{R}_2}(Y_2)$,
(\ref{eq:fifth}) follows from (\ref{eq:assump}) and by writing the
mutual information as the difference between two entropies,
(\ref{eq:six}) is since $h(X^-+J+N_2|X^+)\leq h(X^-+J+N_2)$ and
$h(Y_2|X)=h(J+N_2)$, (\ref{eq:seven}) is by noticing that
$h(J+N_2)=h(X^-+J+N_2|X^-)$ and simply writing the difference
between the entropies as mutual information, (\ref{eq:eigth}) is
because $\mathrm{E}|X^-|^2\leq P_J$ and finally (\ref{eq:final}) is
since $H(V_1,V_2)\leq C_1+C_2$ and $I(\boldsymbol{X}^+;V_1,V_2)\leq
H(\boldsymbol{X}^+)$, where $\mathrm{E}|\boldsymbol{X}^+|^2\leq
\frac{P_X}{P_J}$, and using Theorem 9.7.1 from \cite{CoverThomas}.

\hfill{\QED}

\subsection{Proofs for the Achievable Rate}
\emph{Proof for Cases B and C (Propositions \ref{prop:R_SI_B} and
\ref{prop:I_B}):}\\
Here we avoid reconstructing the whole $J$ at the destination by
utilizing a lattice code and reducing the signals into its Voronoi
cell by a modulo operation. Our scheme is an adaptation of the MLAN
channel technique from \cite{ErezZamir2004}.

First define the lattice code $\mathbb{C}_2$ which is a good source
$P_X$-code, which means that it satisfies, for any $\varepsilon>0$
and adequately large lattice dimension $n$
\begin{equation}\begin{split}\label{eq:good_source}
\log(2\pi e G(\mathbb{C}_2))<\varepsilon\\
P_X=\frac{\int_{\nu_2}||\mathbf{x}||^2d\mathbf{x}}{V_o n}.
\end{split}
\end{equation}
Where $G(\mathbb{C}_2)$, $\nu_2$ and $V_o$ are the normalized second
moment, the Voronoi cell and the Voronoi cell volume of the lattice
associated with $\mathbb{C}_2$, respectively. Such codes are known
to exist \cite{Poltyrev1994}.
\begin{enumerate}
\item \textit{Transmission Scheme}: Transmit the information $W$ as a codeword from a codebook, where every codeword in this codebook is randomly and independently generated by dividing it into many (multi-letter) entries, each generated uniformly i.i.d.
over the Voronoi region of $\mathbb{C}_2$, $\nu_2$. Define the
transmitted codeword as $V$. Add a pseudo random dithering $-U$,
which is uniformly generated over $\nu_2$ and known to all parties,
to get: $X=V-U \mod \mathbb{C}_2$ (modulo Voronoi region of
$\mathbb{C}_2$). This dither is required for the analysis, to ensure
independence of the modulo noise with respect to the message index.
\item \textit{Relaying Scheme}:
Both relays $\mathcal{R}_1$ and $\mathcal{R}_2$ multiply the
received signals by $\alpha>0$, apply $\mod\mathbb{C}_2$, and
quantize the received signal using standard information theory
techniques into $W_1=\alpha Y_1 \mod\mathbb{C}_2+D_1$ and $W_2=\alpha Y_2 \mod\mathbb{C}_2+D_2$. The quantization is given in Appendix \ref{app:Comp}, where $U$,$Y$ in Appendix \ref{app:Comp} are $W_1=\alpha Y_1 \mod\mathbb{C}_2+D_1$ and $\alpha Y_1 \mod\mathbb{C}_2$, respectively. The underlying single letter distortions $D_1$ and
$D_2$ in $W_1,W_2$ are Gaussian with zero mean and are independent with any
other random variable.

A Slepian Wolf encoding is then used on the two \textit{vector} quantized signals,
before transmission to the destination.

\item \textit{Decoding at Destination}:
Now the destination decodes $W_1$ and $W_2$ and calculates
$W_1-W_2$. From the result, it further subtracts the known pseudo
random dither $U$, and applies again modulo $\mathbb{C}_2$ (see
equation (\ref{eq:subsub})). It then finds the vector $\hat{V}$
which is jointly typical with the resulting outcomes of the modulo
operation. The decoded message is the corresponding message index
$W$, if decoding is successful.
\item \textit{Analysis of Performance}:
The independent distortion variance $P_{D_1}$ corresponds to what
is promised by the rate distortion function for any random
variable with variance of $P_X$, and in particular, to $\alpha Y_1
\mod \mathbb{C}_2$ (See Appendix \ref{app:Comp} for the complete proof). The
rate for independent distortion for $Y_1$ is
\begin{equation}\label{eq:independent_distortion}
I(W_1;\alpha Y_1
\mod \mathbb{C}_2)=\frac{1}{2} \log \frac{P_X+P_{D_1}}{P_{D_1}}\leq C_1.
\end{equation}
Notice that taking $D_1$ such that $P_{D_1}$ fulfills
(\ref{eq:independent_distortion}) allow us to chose $D_1$ to be
distributed independently of $\alpha Y_1 \mod \mathbb{C}_2$,
regardless of $\alpha$.

 \emph{For Proposition \ref{prop:R_SI_B}:} Depending on
whether $\alpha^2 P_J< P_X$ or $\alpha^2 P_J>P_X$, using the result in Appendix \ref{app:Comp}, we get for
$P_{D_2}$
\begin{equation}\label{eq:independent_distortion2}
I(W_2;\alpha Y_2
\mod \mathbb{C}_2)=\frac{1}{2} \log \frac{\min\{P_X,\alpha^2 P_J\}}{P_{D_2}}\leq C_2.
\end{equation}
Since $X=V-U$, we can write $\alpha Y_1 - \alpha Y_2 + U= \alpha X
+ \alpha N_1 + U= X+U - (1-\alpha)X + \alpha N_1 = V - (1-\alpha)X
+ \alpha N_1 $, and the following equalities hold
\begin{eqnarray}
Y&=&W_1-W_2 + U\mod \mathbb{C}_2\label{eq:subsub}\\
 &=& \alpha Y_1 + D_1 - \alpha Y_2 - D_2 + U\mod \mathbb{C}_2\\
 &=& V - (1-\alpha)X + \alpha N_1 + D_1 - D_2\mod \mathbb{C}_2.\label{eq:inter_sum}
\end{eqnarray}
Define
\begin{equation}\label{eq:equiv_noise}
N_{eq} =  \alpha N_1+D_1-D_2 - (1-\alpha)X.
\end{equation}
with
\begin{equation}\label{eq:noise}
P_{N_{eq}}=\alpha^2 P_{N_1}+ (1-\alpha)^2P_X + P_{D_1}+P_{D_2}.
\end{equation}
Encoding according to $V$ which is uniformly distributed over
$\mathbb{C}_2$ gives an achievable rate of (See Inflated Lattice
Lemma in \cite{ErezZamir2004})
\begin{equation}\label{eq:achievable_R2}
R>\frac{1}{n}\left(\log_2\frac{V_o}{G(\mathbb{C}_2)}-H(N_{eq})\right)-\delta=\frac{1}{2}\log_2(2\pi
e P_X)-\frac{1}{2}\log_2\left(2\pi e P_{N_{eq}}\right)-\delta.
\end{equation}
Setting $\alpha$ to maximize the achievable rate
\begin{equation}\label{eq:alpha}
\alpha=\frac{P_X}{P_X+P_{N_1}},
\end{equation}
along with (\ref{eq:independent_distortion}) and
(\ref{eq:independent_distortion2}) results in
\begin{equation}\label{eq:R_2_stage3}
R>\frac{1}{2}\log_2\left(\frac{P_X}{\frac{P_X P_{N_1}}{P_X+
P_{N_1}}+P_{D_1}+P_{D_2}}\right)-\delta=\frac{1}{2}\log_2\left(\frac{P_X}{\frac{P_X
P_{N_1}}{P_X+
P_{N_1}}+\frac{P_X}{2^{2C_1}-1}+\min\{P_X,P_J\frac{P^2_X}{(P_X+P_{N_1})^2}\}2^{-2C_2}}\right)-\delta.
\end{equation}
Considering that $P_{N_1}=1$, after some simple algebra,
(\ref{eq:R_2_stage3}) becomes (\ref{eq:R_2_stage4}). \hfill{\QED}
\end{enumerate}
\emph{For Proposition \ref{prop:I_B}:} On one hand
\begin{equation}\label{eq:independent_distortion3_1}
\frac{1}{n}H(\bold{W_2}|\bold{W_1})\leq
\frac{1}{n}H(\bold{W_2})\leq \frac{1}{2} \log
\left(1+\frac{P_X}{P_{D_2}}\right),
\end{equation}
on the other hand
\begin{equation}\label{eq:independent_distortion3_2}
\frac{1}{n}H(\bold{W_2}|\bold{W_1})=\frac{1}{n}H(\bold{W_1}+\bold{W_2}|\bold{W_1})\leq
\frac{1}{n}H(\bold{W_1}+\bold{W_2})\leq\frac{1}{2} \log
\left(1+\frac{\alpha^2(4P_J+P_{N_1}+P_{N_2})+P_{D_1}}{P_{D_2}}\right).
\end{equation}
So for a successful Slepian-Wolf decoding we require that the
minimum between the right hand sides of equations
(\ref{eq:independent_distortion3_1}) and
(\ref{eq:independent_distortion3_2}) be smaller than $C_2$. This
brings us to
\begin{equation}\label{eq:independent_distortion3}
\frac{1}{2} \log
\left(1+\frac{\min\{P_X,\alpha^2(4P_J+P_{N_1}+P_{N_2})+P_{D_1}\}}{P_{D_2}}\right)\leq
C_2.
\end{equation}
As in (\ref{eq:inter_sum}), here we have
\begin{equation}
Y= V - (1-2\alpha)X + \alpha (N_1+N_2) + D_1 + D_2\mod
\mathbb{C}_2.\label{eq:inter_sum2}
\end{equation}
and $P_{N_{eq}}$ in this case, is
\begin{equation}\label{eq:noise2}
P_{N_{eq}}=\alpha^2 (P_{N_1}+P_{N_2})+ (1-2\alpha)^2P_X +
P_{D_1}+P_{D_2}.
\end{equation}
Taking
\begin{equation}\label{eq:alpha2}
\alpha=\frac{2P_X}{4P_X+P_{N_1}+P_{N_2}},
\end{equation}
we get (considering that $P_{N_1}=P_{N_2}=1$)
\begin{equation}\label{eq:R_2_stage32}
R>\frac{1}{2}\log_2\left(\frac{P_X}{\frac{P_X
(P_{N_1}+P_{N_2})}{4P_X+
P_{N_1}+P_{N_2}}+P_{D_1}+P_{D_2}}\right)-\delta\geq\frac{1}{2}\log_2\left(\frac{P_X}{\frac{1}{2}+\frac{P_X}{2^{2C_1}-1}+\frac{\min\{P_X,4P_J+2+\frac{P_X}{2^{2C_1}-1}\}}{2^{2C_2}-1}}\right)-\delta.
\end{equation}

The proof can be further replicated also when interchanging the
indices 1 and 2.
 \hfill{\QED}

\section{Conclusions and Discussions}
In this paper, we derive both inner and outer bounds of the
communication rate, for three common distributed reception
scenarios, with unknown interference. The three scenarios characterize the
very low noise case of the more general case of
distributed reception of wanted signal plus unknown interference. The inner bounds rely on
lattice coding, since standard random coding techniques do not
provide satisfactory results, in general. Outer bounds based on
the cut-set technique are derived, and additional tighter bound is
derived by using multi-letter techniques, for a case where the
cut-set bound does not suffice. This case includes two relays,
which receive both the desired signal and the interference. The
generally loose inner and outer bounds which coincide at
asymptotically large powers of the transmitter and the interferer,
are used to derive the scaling laws. These scaling laws reveal
that in order to overcome interference, a defined amount of
information about the interference must be known at the
destination. The proposed scheme for the inner bound, is also
robust against the interference statistics, code, modulation etc.\\
The model is intimately related to the case of two independent transmitters. Then the transmission of one transmitter can be treated
as interference with power $P_J$ as in this paper. This approach is beneficial when the rate in which the interfering transmitter $R_J$ is high, so codebook knowledge is useless. If in addition the power of the interfering transmitter is high $P_J>P_X$, then the achievable rates in this paper provide a better approach than the standard compress-and-forward.
\section*{Acknowledgment}
This research was supported by the NEWCOM++ network of excellence
and the ISRC Consortium.
\appendices
\section{Proofs for the asymptotic gaps}\label{app:gap}
Define the gap between the achievable rate and the outer bound as $\Delta$.
\subsection{Case A}
For Case A, where $C_2=\frac{1}{2}\log_2(1+P_X)$ and $P_J>P_X$ we get
\begin{multline}
\Delta=C_2 + \frac{1}{2}\log_2\left(1+\frac{P_X}{P_J+1}\right) - \frac{1}{2}\log_2
\bigg(\frac{1+P_X}{1+\min\left\{1+P_X,\frac{P_JP_X}{P_X+1}\right\}2^{-2C_2}}
\bigg)\leq\\ \frac{1}{2}\log_2\left(1+\frac{P_X}{P_J+1}\right) + \frac{1}{2}\log_2(2)\leq \frac{1}{2}\log_2(1.5\times2)=0.7925,
\end{multline}
whereas for $P_J<P_X$ and $C_2=\frac{1}{2}\log_2(P_J)$ we get
\begin{multline}
\Delta= C_2 + \frac{1}{2}\log_2\left(1+\frac{P_X}{P_J+1}\right) - \frac{1}{2}\log_2
\bigg(\frac{1+P_X}{1+\min\left\{1+P_X,\frac{P_JP_X}{P_X+1}\right\}2^{-2C_2}}
\bigg)\leq\\ \frac{1}{2}\log_2\left(P_J+P_X\right) - \frac{1}{2}\log_2
\bigg(\frac{1+P_X}{1+\frac{P_X}{P_X+1}}\bigg)\leq \frac{1}{2}\log_2(2*2)=1.
\end{multline}
So overall for Case A, $\Delta\leq 1$.
\subsection{Case B}
For Case B, where $C_1=\frac{1}{2}\log_2(1+P_X)$,$C_2=\frac{1}{2}\log_2(P_J)$ and $P_X>1$, we get that
\begin{equation}
\min\left\{1+P_X,\frac{P_JP_X}{P_X+1}\right\}2^{-2C_2}(2^{2C_1}-1)\leq 1+P_X.
\end{equation}
Which gives
\begin{multline}
\Delta=C_2+\frac{1}{2}\log_2\left(1+\frac{P_X}{P_J+1}\right) - \frac{1}{2}\log_2\left(\frac{(1+P_X)(2^{2C_1}-1)}{P_X+
2^{2C_1}+\min\left\{1+P_X,\frac{P_JP_X}{P_X+1}\right\}2^{-2C_2}(2^{2C_1}-1)}\right)\leq\\
\frac{1}{2}\log_2\left(P_J+P_X\right)-\frac{1}{2}\log_2\left(\frac{(1+P_X)P_X}{P_X+(1+P_X)+(1+P_X)}\right)\leq\\
\frac{1}{2}\log_2(2)+\frac{1}{2}\log_2\left(P_X\right)-\frac{1}{2}\log_2\left(\frac{P_X}{3}\right)\leq 1.29.
\end{multline}
So overall for Case B, $\Delta\leq 1.29$.
\subsection{Case C}
For Case C, where $P_J<P_X<\frac{(1+P_X)^2}{P_X}$, the modulo upper bound is relevant. So we have
\begin{multline}
\Delta=\frac{1}{2}\left(C_1+C_2+\frac{1}{2}\log_2\left(1+\frac{P_X}{P_J}\right)\right)+\frac{1}{4}\log_2(8\pi
e)\\-\frac{1}{2}\log_2\left(\frac{P_X}{\frac{P_X
}{P_X+
1}+\frac{P_X}{2^{2C_1}-1}+P_J\frac{P_X}{(P_X+1)^2}2^{-2C_2}}\right).
\end{multline}
We evaluate the gap for corner case where we take $C_1=\frac{1}{2}\log_2(1+P_X)$ and $C_2=\frac{1}{2}\log_2(P_J)$.
\begin{multline}\label{eq:first_gap}
\Delta=\frac{1}{2}\left(\frac{1}{2}\log_2(1+P_X)+\frac{1}{2}\log_2(P_J)+\frac{1}{2}\log_2\left(1+\frac{P_X}{P_J}\right)\right)+\frac{1}{4}\log_2(8\pi
e)\\-\frac{1}{2}\log_2\left(\frac{P_X}{\frac{P_X}{P_X+
1}+\frac{P_X}{P_X}+\frac{P_X}{(P_X+1)^2}}\right).
\end{multline}
Since $\frac{P_X}{P_X+1}+1+\frac{P_X}{(P_X+1)^2}=\frac{P_X(1+P_X)+(1+P_X)^2+P_X}{(P_X+1)^2}\leq 3$, (\ref{eq:first_gap}) using $P_J>1$ gives
\begin{equation}
\Delta\leq\frac{1}{4}\log_2(4)+\frac{1}{2}\log_2(P_X)+\frac{1}{4}\log_2(8\pi
e)-\frac{1}{2}\log_2\left(\frac{P_X}{3}\right)=\frac{1}{4}\log_2(8\times4\times9\pi
e)=2.816.
\end{equation}
For $P_J>\frac{(1+P_X)^2}{P_X}$, the relevant upper bound is the cut-set bound. We find the gap for $C_1=C_2=\frac{1}{2}\log_2(1+P_X)$, which gives the correct scaling. So here we have (also using $P_X>1$):
\begin{multline}
\Delta=C_1+\frac{1}{2}\log_2\left(1+\frac{P_X}{P_J+1}\right)-\frac{1}{2}\log_2\left(\frac{P_X}{\frac{P_X
}{P_X+
1}+\frac{P_X}{2^{2C_1}-1}+P_X2^{-2C_2}}\right)\leq\\C_1 + \frac{1}{2}\log_2\left(2\right)-\frac{1}{2}\log_2\left(\frac{P_X}{\frac{P_X
}{P_X+
1}+1+\frac{P_X}{1+P_X}}\right)\leq \frac{1}{2}\log_2(2\times4)=1.5.
\end{multline}
So overall, in the low noise power limit, when $R\sim 0.5$, for cases A,B and C the gap between the achievable rate and the outer bound is bounded to 1,1.29 and 2.816 bits, respectively.\\
\section{Proof for scaling laws of Case C}\label{app:scaling}
\begin{proof}\\
\emph{Necessary conditions:} The outer bound in
(\ref{eq:cutset_C}),(\ref{eq:R_imprv_B}) is written as a rate region
for $C_1,C_2$ in (\ref{eq:scaling_C}), such that four constraints
are met, where two constraints limit $C_1+C_2$ and the other two
constraints limit $C_1$ and $C_2$ separately.\\
\emph{Sufficient conditions:} The outer bound (\ref{eq:scaling_C})
consists of three inequalities, which leads to two intersections points (see Figure \ref{fig:region}).
Thus the
entire region is achievable, for example by using time sharing,
provided the point where the capacities of the links are $C_1\sim
\max\left\{R,\frac{1}{2}\log_2\left(1+\frac{P_X}{P_J}\right)\right\},C_2\sim
R-\frac{1}{2}\log_2\left(1+\frac{P_X}{P_J}\right)$ (P1 in Figure
\ref{fig:region}) corresponds to a scheme with the same scaling of
the reliable rate as $R$. The proof is then completed by repeating
the same arguments for the second point (P2 in Figure
\ref{fig:region}).

In case
$R\lesssim\frac{1}{2}\log_2\left(1+\frac{P_X}{P_J+1}\right)$, use
$P_X$ to transmit so that the message will be separately decoded
at the agents, where $C_1=R$ and $C_2=0$. Since the agents receive
the transmitted signal with signal to noise plus interference
ratio of $\frac{P_X}{P_J+1}$, decoding is reliable. In case
$R\gtrsim\frac{1}{2}\log_2\left(1+\frac{P_X}{P_J}\right)$, use the
scheme from Proposition \ref{prop:I_B}, which achieves the rate of
(\ref{eq:achiv_C}), with $C_1\sim R$ and $C_2\sim
R-\frac{1}{2}\log_2\left(1+\frac{P_X}{P_J}\right)$. This rate is
\begin{equation}\label{eq:achiv_C22}
\frac{1}{2}\log_2\left(\frac{P_X}{P_X 2^{-2C_1}+\min\{P_X,P_J\}2^{-2C_2}}\right)=\frac{1}{2}\log_2\left(2^{2R}\frac{P_X}{P_X +\min\{P_X,P_J\}(1+\frac{P_X}{P_J})}\right)\sim R.
\end{equation}
The bounded gap between the achievable rate and the upper bound is evaluated in Appendix \ref{app:gap}.
\end{proof}

\section{Proof for compression}\label{app:Comp}
Proof that $R>I(Y;U)$ is sufficient for the relay which received $Y$ to forward $U$ to the final destination.
For any $\epsilon>0$,
\begin{enumerate}
\item \textit{Preliminaries} As is commonly done (see \cite{CoverThomas}, section 13.6), define
the $\eps$-typical set $\mathbf{T}_\eps$ of vectors $\boldsymbol{a}_{1,2}$, with relation to the probability density function $P_{a_{1,2}}$ as
\begin{equation}\label{eq:typical_set_def}
\mathbf{T}_\eps\defeq\ \Bigg\{\boldsymbol{a}_{1,2}:\ \forall
\mathcal{S}\subseteq\{1,2\},\
\left|-\frac{1}{n}\log_2\left(P^n_{a_\mathcal{S}}(\boldsymbol{a}_\mathcal{S})\right)-h(P_{a_\mathcal{S}})\right|<
\eps,\Bigg\}
\end{equation}
where $P^n_{a_\mathcal{S}}(\boldsymbol{a}_\mathcal{S})=\prod_{i=1}^n P_{a_\mathcal{S}}\left((\boldsymbol{a}_\mathcal{S})_i\right)$, and $h(P_{a_\mathcal{S}})$ is the differential entropy of the probability density function $P_{a_\mathcal{S}}$, where $\mathcal{S}=[1,2,\{1,2\}]$.

\begin{lem}\label{lem:AEP}(AEP)
For any $\eps>0$, there exist $n^*$ such that for all $n>n^*$ and
$\boldsymbol{a}_{1,2}\sim\prod P_{A_{1,2}}$ we have
\begin{equation}
P(\boldsymbol{a}_{1,2} \in \mathbf{T}_\eps)\geq 1-\eps.
\end{equation}
\begin{proof} See \cite{CoverThomas} Theorem 9.2.2. \end{proof}
\end{lem}

\begin{lem}\label{lem:jointly_typical}
Let $\boldsymbol{a}_{1,2}$ be generated according to
\begin{equation}\label{eq:true_probability}
\boldsymbol{a}_{1,2}\sim \prod_{i=1}^n P_{a_1}((\mathbf{a}_1)_i)P_{a_2}((\mathbf{a}_2)_i).
\end{equation}
Then we have
\begin{equation}
\Pr(\boldsymbol{a}_{1,2}\in\mathbf{T}_\eps)=\frac{\Pr(\boldsymbol{a}_{1,2}\in\mathbf{T}_\eps\bigcup \boldsymbol{a}_{1}\in\mathbf{T}_\eps\bigcup \boldsymbol{a}_{2}\in\mathbf{T}_\eps)}{\Pr(\boldsymbol{a}_{1}\in\mathbf{T}_\eps)\Pr(\boldsymbol{a}_{2}\in\mathbf{T}_\eps)}\geq 
2^{-n[h(a_1)+h(a_2)-h(a_{1,2})+\eps_1]}=2^{-n[I(a_1;a_2)+\eps_1]}
\end{equation}\vspace{-0.3in}
where $\eps_1\rightarrow 0$ as $\eps\rightarrow 0$.\\
\end{lem}

\item \textit{Code generation} Randomly generate $2^{nI(Y;U)}$ codewords $\mathbf{U}$, according to i.i.d. distribution
$\Pi_{i=1}^{n}P_U(U_i)$. Index these codewords by $z\in[1, 2^{nI(Y;U)}]$. The codebook is made available to the relay and the destination.
\item \textit{Compression} After receiving the vector $\mathbf{Y}$, the relay searches for $z$ such that $\{\mathbf{U}(z),\mathbf{Y}\}\in\mathbf{T}_\epsilon$. If no such $z$ is found, the relay sends $z=1$.
\item \textit{Error Analysis} The probability of two independent random variables $\mathbf{U},\mathbf{Y}$ to be jointly typical is lower bounded by
\begin{equation}
2^{-n[I(Y;U)+\epsilon]}.
\end{equation}
Thus the probability that no such $z$ is jointly typical is upper bounded by
\begin{equation}
\left(1-2^{-n[I(Y;U)+\epsilon]}\right)^{2^{nR}},
\end{equation}
which tend to zero as $n$ gets large as long as $R>I(Y;U)+\epsilon$. \QED
\end{enumerate}

\end{document}